# A Zonal Similarity Analysis of Velocity Profiles in Wall-Bounded Turbulent Shear Flows

**Trinh, Khanh Tuoc**
Institute Of Food Nutrition And Human Health
Massey University, New Zealand
K.T.Trinh@massey.ac.nz

## Abstract

It is argued that there are three distinct zones in a wall bounded turbulent flow field dominated by three completely different mechanisms:

- An outer region where the velocity profile is determined by the pressure distribution
- A highly active wall layer dominated by a sequence of inrush-sweep and ejections, and
- An intermediate region well described by the traditional logarithmic law proposed by independently Millikan and Prandtl. The log-law and the wall layer are sometimes referred to as the inner region. Under these conditions, a unique set of normalisation parameters cannot possibly apply to all three zones. The inner region can be more successfully represented by normalising the distance and velocity with the values of these scales at the edge of the wall layer since they are shared by both the wall layer and the log-law region.

The application of this similarity analysis has successfully collapsed extensive published data for the inner region covering a range of Reynolds numbers from 3000 to 1,000,000 in a variety of geometries including cylindrical pipes, external boundary layers on flat plates, recirculation regions behind a sudden channel expansion, converging rectangular channels and oscillating pipe flows into a unique curve. The normalisation also collapsed drag reducing flow involving ribblets, power law fluids and viscoelastic fluids onto the standard Newtonian curve.

## 1 Introduction

In the search for scaling laws for wall-bounded turbulent flow, it is customary to divide the velocity field into two regions: an inner region close to the wall and an outer region, the turbulent core, e.g. Panton (1990), . It is much easier to obtain simplified models for each of these regions than to solve the original Navier-Stokes equations. These asymptotic solutions are then matched to give a description of the whole flow field.

Prandtl (1935) made the first attempt at a similar representation of the turbulent velocity profile by normalisation with the wall parameters: the friction velocity $u_* = \sqrt{\tau_w/\rho}$ and the kinematic viscosity $\nu$. Prandtl expected the resulting plot to be independent of the Reynolds numbers and called it the universal velocity profile. From his mixing-length theory, Prandtl derived the semi-logarithmic law of the wall, henceforth abbreviated to the log-law:

$$U^+ = A \,\mathrm{Ln}\, y^+ + B \qquad (1)$$

Nikuradse (1932) measured the velocity profiles in turbulent pipe flow and obtained empirically A = 2.5, B = 5.5 for Re > 6,000. However, even these early experiments showed a small but definite Reynolds number effect on the universal velocity profile.

The log-law does not apply very close to the wall where Prandtl reasoned that the velocity fluctuations must be damped and viscous flow must prevail. While the log-law is still widely used in turbulent flow simulations, the physical arguments of the mixing-length proposed by Prandtl have been largely discredited. Millikan (1939) argued that the outer region can be normalised in terms of the outer variables, the velocity at the pipe axis or edge of the boundary later $U_m$ and the radius R or the thickness $\delta$ of the boundary layer, and the inner region with the wall parameters. By matching these two asymptotic solutions, Millikan showed that the intermediate region must obey a log-law, which is therefore not dependent on the detailed physical assumptions of Prandtl's mixing-length theory. An alternative to the log-law is a power law first analysed in detail by Nikuradse (1933) based on the success of the empirical Blasius friction factor correlation (1913)

The solution of the Navier-Stokes equations in the outer region is greatly influenced by the pressure term, which is itself defined by the geometry of the flow field. Launder and Spalding (1974) were the first to take advantage of advances in computer technology to develop numerical methods for analysis of complex multi-dimensional flow systems. A great number of models have been proposed for flow simulation but they only apply to the outer region and must be linked to the wall conditions by empirical wall functions. Bradshaw, Launder et al.(1991) noted that the result of the simulations were quite insensitive to the models used for the outer region but succeeded in giving a reasonable description of the turbulent flow field whenever the codes used the log-law to connect with the wall.

There has been a resurgence of interest in scaling laws in recent years. Many authors have proposed high-order corrections to the classical the log-law (Tennekes, 1966, Azfal and Yanik, 1973, Wosnik et al., 2000). Zagarola and Smits (1998) attributed the lack of consensus on the scaling in the intermediate region (which they call overlap) or even on its existence partly to the lack of adequate experimental data and obtained detailed measurements of pipe flow velocity profiles over a wide range of Reynolds numbers. Their "Superpipe" data exceeded the maximum Reynolds number of even the data of Nikuradse. They used an equation of the form

$$U^+ = \frac{1}{\kappa}\ln(y^+ + a^+) + B \tag{2}$$

The concept of an empirical shift parameter $a^+$ dates back a long time to Duncan et al. (1960) but was largely ignored until Oberlack (1999) justified it with Lie-group analysis and Wosnik, et al. (2000) suggested that it was required to account for a mesolayer, a region in which the overlap argument of the mean velocity profiles in the outer and wall regions exist but where the separation between the energy and dissipation scales is not large enough for inertially dominated turbulence to exist. Even more complex correlations have been proposed by Buschmann and Gad-el-Hak (2004, 2003) who included higher order terms in $y^+$ and $R^+$ in the classical overlap analysis. Interestingly, these latter authors argued that the Karman constant $\kappa$ must be modified by a factor which is itself a function of the Reynolds number, an approach taken by McKeon et al. (2004) for a more complex derivation of the friction factor–Reynolds number relationship that fitted better the Princeton Superpipe data than the classical Prandtl-Karman formula.

A different approach was taken by Fife et al. (2005) who argued that the regions of pressure driven turbulent channel flows are characterised by a balance between the gradients of the viscous and Reynolds stresses (stress gradient balance). They argued that each of these regions is characterised by an intrinsic hierarchy of “scaling layers” which can be defined in terms of a function $A(y^+)$ which under certain conditions can be related to the Karman constant $\kappa$ but remain somewhat non-committal about the physical significance of A and the exact situations under which it is, like Kaman’s $\kappa$ a constant and not a function.

In this contribution, a scaling method is proposed, which removes the Reynolds number effect to give true similarity profiles for the inner region for both Newtonian and non-Newtonian fluids in a variety of flow configurations.

# 2 Theory

## 2.1 The wall layer and the inner region

In 1967, (Kline et al.) reported their now classic hydrogen bubble visualisation of events near the wall and ushered in a new area of turbulence research based on the so-called coherent structures. Despite the prevalence of viscous diffusion of momentum close to the wall, the flow was not laminar in the steady state sense envisaged by Prandtl. Instead the region near the wall was the most active in the entire flow field. In plan view, Kline *et. al.* observed a typical pattern of alternate low – and high-speed streaks. The low-speed streaks tended to lift, oscillate and eventually eject away from the wall in a violent burst. In side view, they recorded periodic inrushes of fast fluid from the outer region towards the wall followed by a vortical sweep along the wall. The low-speed streaks appeared to be made up of fluid underneath the travelling vortex as shown in Figure 1.

The bursts can be compared to jets of fluids that penetrate into the main flow, and get slowly deflected until they become eventually aligned with the direction of the main flow.

Figure 1. Visualisation of a cycle of the wall layer process drawn after the observations of Kline et al.(1967) and regions in the flow field.

Walker (1978) was the first to show that a vortex travelling above the wall induces a viscous sub-boundary layer underneath its path. His work has been extensively verified with simulations of vortices of different configurations. Perridier et al. (1991), for example, have shown how this viscous sub-boundary layer eventually erupts in an explosive event they call viscous-inviscid interaction. Indeed Suponistky et al. (2005) show that a vortex of any shape travelling along the proximity of a wall can eventually transform into a hairpin vortex. The low-speed-streak phase is much more persistent than the ejection phase and dominates their relative contribution to the time-averaged velocity profile Walker et al.(1989). The edge of the wall layer may be defined as the position of maximum penetration of wall retardation in the main flow through diffusion of viscous momentum and coincides with the maximum thickness $\delta_v$ of this induced viscous sub-boundary layer. This occurs at the point of bursting, or ejection. Thus the inner region can be further divided into a wall layer and a log-law zone.

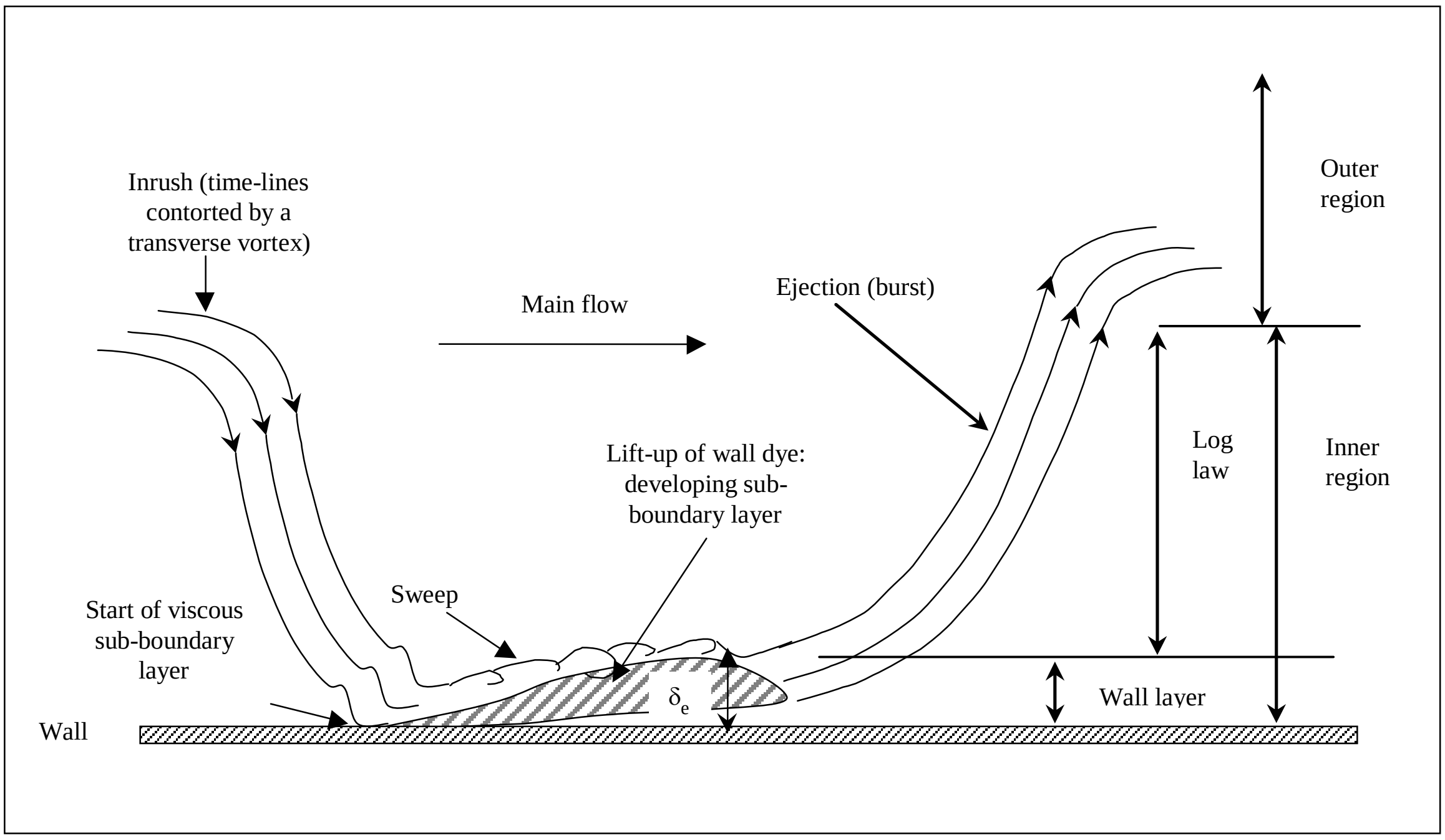

It is clear that the time averaged velocity distributions in the wall layer and the outer region arise from very different transient events and cannot be described by the same scaling laws. It is proposed that neither the wall parameters nor the outer variables should be used. Since the wall layer and the log-law zones are adjacent it is proposed that the correct normalising variables for these two zones are those found at their interface: the velocity $U_v$ at the edge of the wall layer and its thickness $\delta_v$.

## 2.2 Determination of the scaling parameters

The scaling parameters $U_v$ and $\delta_v$ can be determined from the measured velocity profiles to be normalised by one of several methods (Trinh, 1992b).

**Method 1: Slope of logarithmic law**

The region immediately outside the wall-layer obeys the log-law and the velocity profile plotted on log-normal coordinates can be fitted with a straight line. The point of departure of the measured velocity profile from this straight line is taken as the edge of the wall layer (see Figure 2a). This method is conceptually correct but not very sensitive because the change in slope of measured velocity profiles near the edge of the wall layer is slow. Karman (1934), for example, has suggested that the log-law may be applied to a position down to $y^+ = 30$ called the edge of the buffer layer, much closer to the wall than proposed by Schlichting (1960) who quotes a wall layer thickness $\delta_\nu^+ \approx 70$. This relatively poor resolution is evident in Figure 2a.

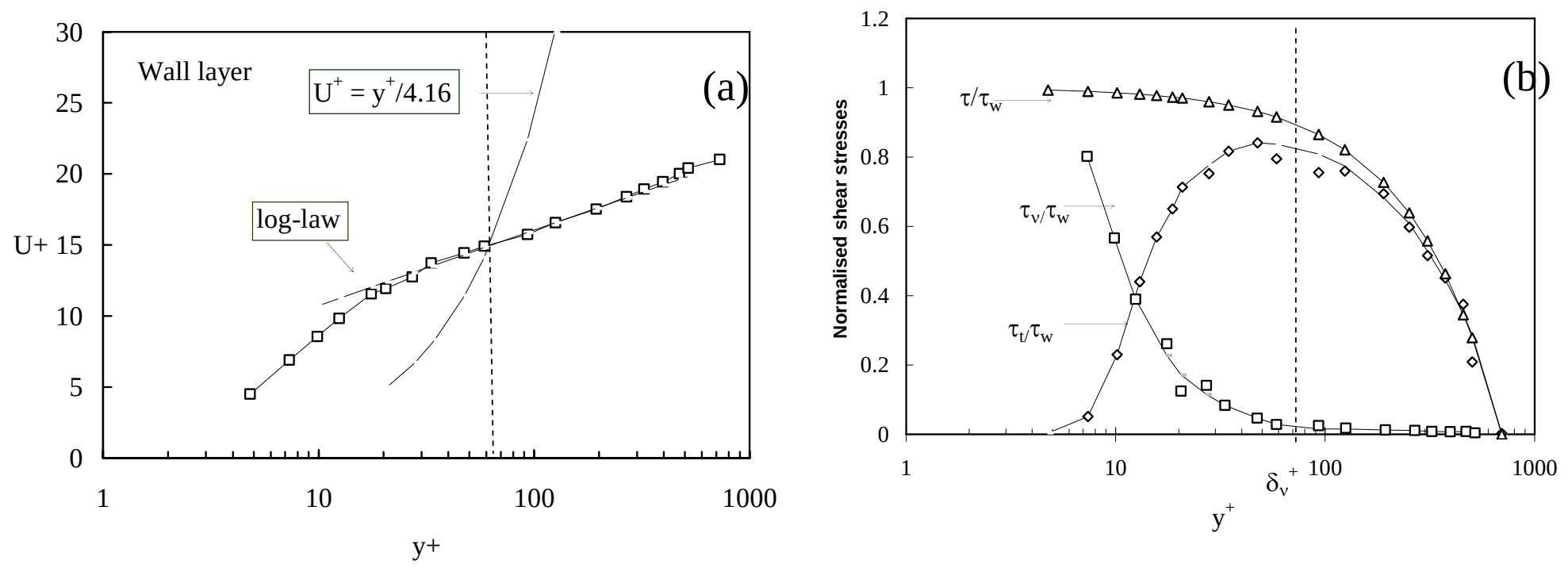

Figure 2. Edge of wall layer determined (a) from slope of log-law and by intersection with the Stokes solution, (b) from normalised viscous stress distribution. Data of Wei and Wilmarth (1989), Re = 14914.

**Method 2: The distribution of shear stresses**

A more direct and fundamental method is based on the distribution of the viscous stresses. In pipe flow, the shear stress $\tau$ varies linearly with distance from the wall (Schlichting, 1960) and is made up of a viscous contribution $\tau_\nu$ and an eddy or turbulent contribution $\tau_t$. These can be calculated from the velocity profile and shear stress at the wall $\tau_w$:

$$\frac{\tau_\nu}{\tau_w} = \frac{\tau_\nu}{\tau}(1 - y^+ / R^+) = \frac{dU^+}{dy^+} \tag{3}$$

At the edge of the wall layer, the local viscous stress must be negligibly small with respect to the turbulent shear stress as shown in Figure 2b. Since the viscous stress decays exponentially with distance $y^+$, an arbitrary cut-off value at the edge of the wall layer is assumed, in this example 4%.

**Method 3: Use of the Stokes solution to model the unsteady viscous sub-layer near the wall**

A number of simplified models have been proposed to describe the intermittent wall layer, e.g. Walker et al. (1989). Einstein and Li (1956) have proposed that we can use the simple Stokes solution (1851) for flow impulsively started over a flat plate. This model was surprisingly successful in describing the time averaged velocity profile over the buffer layer ($0 < y^+ < 30$) and was further developed by others (Hanratty, 1956, Reichardt, 1971, Black, 1969, Meek and Baer, 1970).

The edge of the Stokes layer is defined by (Stokes, 1851)

$$\delta_v^+ = 4.16\, U_v^+ \tag{4}$$

The instantaneous velocity is given by

$$\frac{u}{U_v} = erf\left(\frac{y}{\sqrt{4\nu t}}\right) \tag{5}$$

The parameters $U_v^+$ and $\delta_v^+$ are obtained simply by matching equation (4) with the measured velocity profiles (see Figure 2a).

All three methods give the same estimate of the wall layer thickness as shown in Figures 2a and 2b. This thickness can be determined phenomenologically by methods 1 and 2 without any prior modelling assumption but method 3 has been found to be easier to apply and gives a sharper estimate.

## 3 Results of the similarity analysis

The zonal similarity analysis has been applied to a number of flow situations. These included Newtonian flow in cylindrical pipes (Nikuradse, 1932, Lawn, 1971, Reichardt, 1943, Laufer, 1954, Bogue, 1961, Wei and Willmarth, 1989), between parallel plates (Schlinder and Sage, 1953) , at the bottom of an agitated vessel (Molerus and W., 1987) in boundary layers on a flat plate (Kline et al., 1967, Hoffman and Mohammadi, 1991), in manipulated boundary layers (Bandyopadhyay, 1986), in converging channels (Tanaka and Yabuki, 1986), behind a backward facing step (Devenport and Sutton, 1991) and in oscillating pipe flow (Akhavan et al., 1991), flow of purely viscous power law fluids (Bogue, 1961) and viscoelastic fluids (Wells, 1965, Pinho and Whitelaw, 1990). Some of the most typical flow configurations are shown in Figures 3 to 7.

## 3.1 Flow in cylindrical pipes

The effect of Reynolds number on the normalised velocity profiles is best illustrated by studying flow in cylindrical pipes, which has been widely investigated.

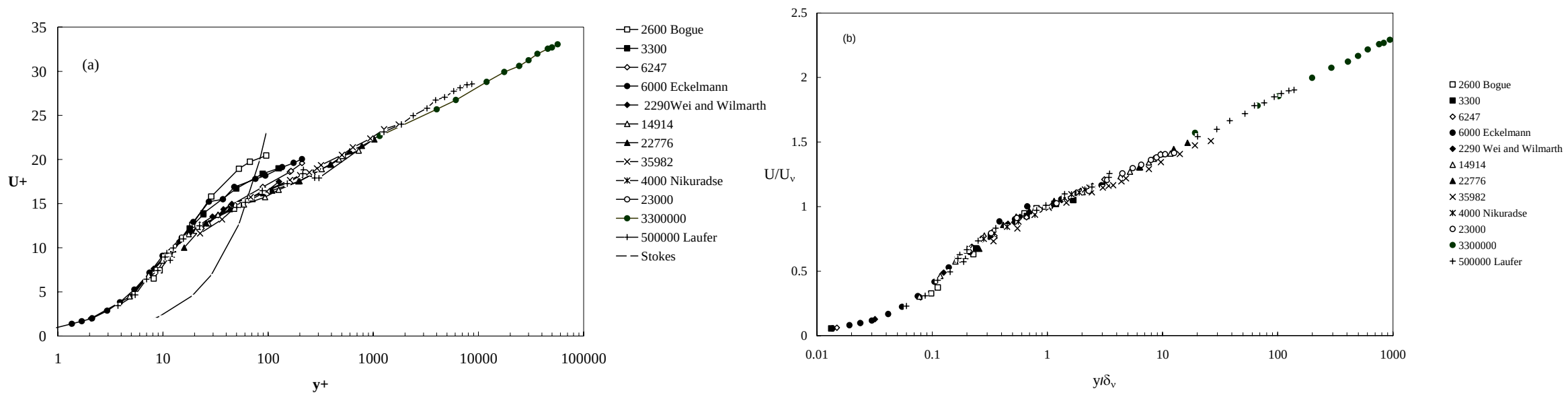

Figure 3. (a) Velocity profiles normalised with the wall parameters and intersection with Stokes' solution (b) Zonal similarity representation. Numbers indicate values of Re. Data from Bogue (1962), Eckelmann (1974), Wei and Wilmarth (1989) Nikuradse (1932) and Laufer (1954)

Figure 3a shows that the edge of the wall layer, determined by the intersection of the Stokes solution with the measured profiles is a function of the Reynolds number. The zonal similarity succeeds in collapsing data from a variety of authors over a range of Reynolds numbers from 2600 to 3,300,000 into a single curve (Figure 3b).

## 3.2 Flow behind a backward facing step

Devenport and Sutton (1991)reported measurements of flow behind a backward facing step both with and without a centre body downstream (Figure 4).

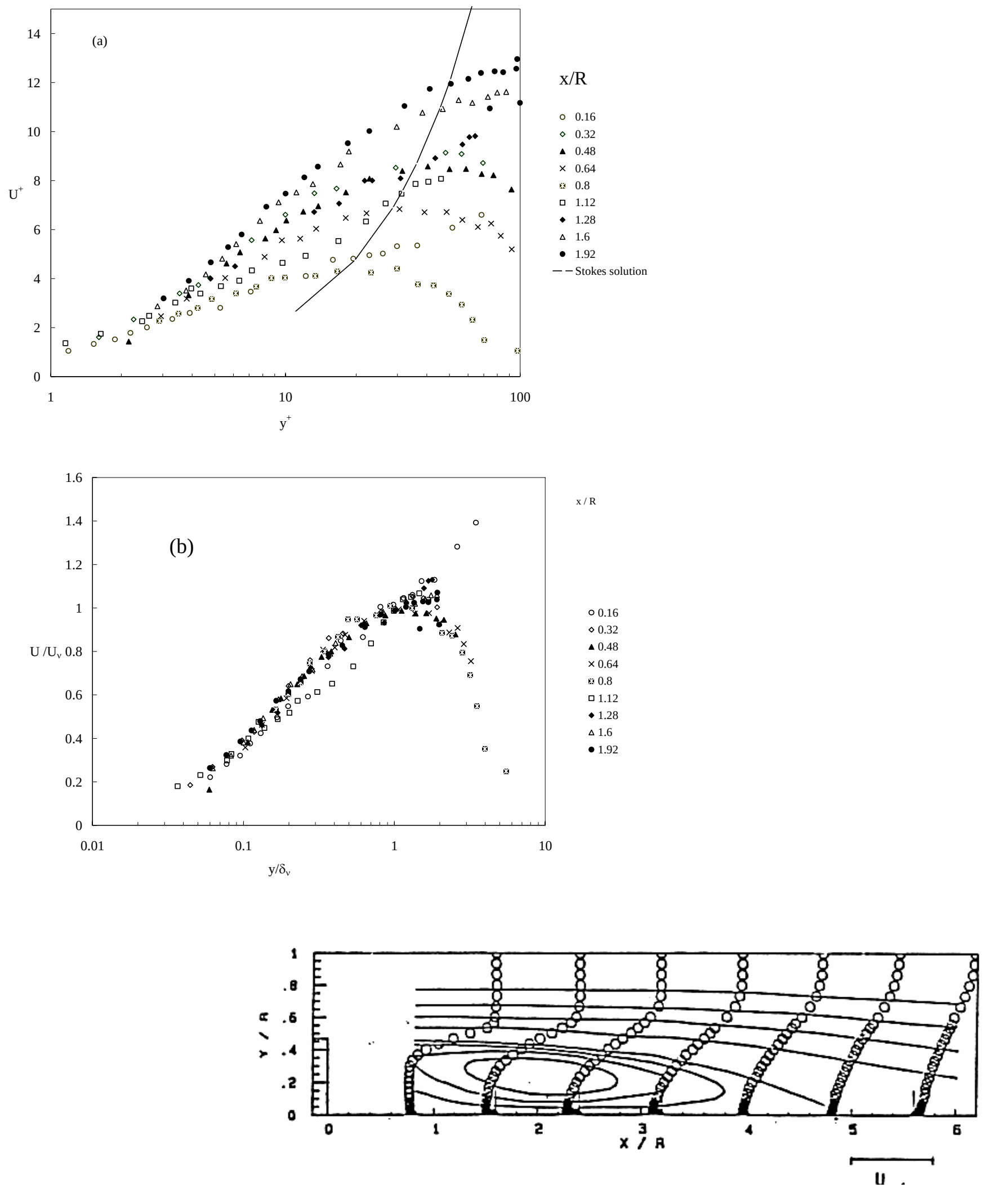

Figure 4. Near-wall mean velocity profiles in separated and reattaching flow behind a backward facing step. (a) normalised with wall parameters (b) Zonal similarity profile (c) Configuration and streamlines. Data of Devenport and Sutton (1991)

This example illustrates the effect of distance from a stagnation point on the velocity profiles. Figure 4c shows the existence of a recirculation region behind the step and Figure 4a shows the velocity distribution at various distances from the point of flow detachment. The zonal similarity profile in Figure 4b shows that the normalised distribution in the wall layer is quite insensitive to the development of the flow pattern outside the recirculation zone

### 3.3 Flow near a flat surface

The effect of pressure on the similar velocity profiles can be illustrated by showing the data for parallel plates (Schlinder and Sage, 1953) where the pressure gradient above the flat wall is, of course, positive and data for flow along flat plates with various pressure gradients (Kline et al., 1967).

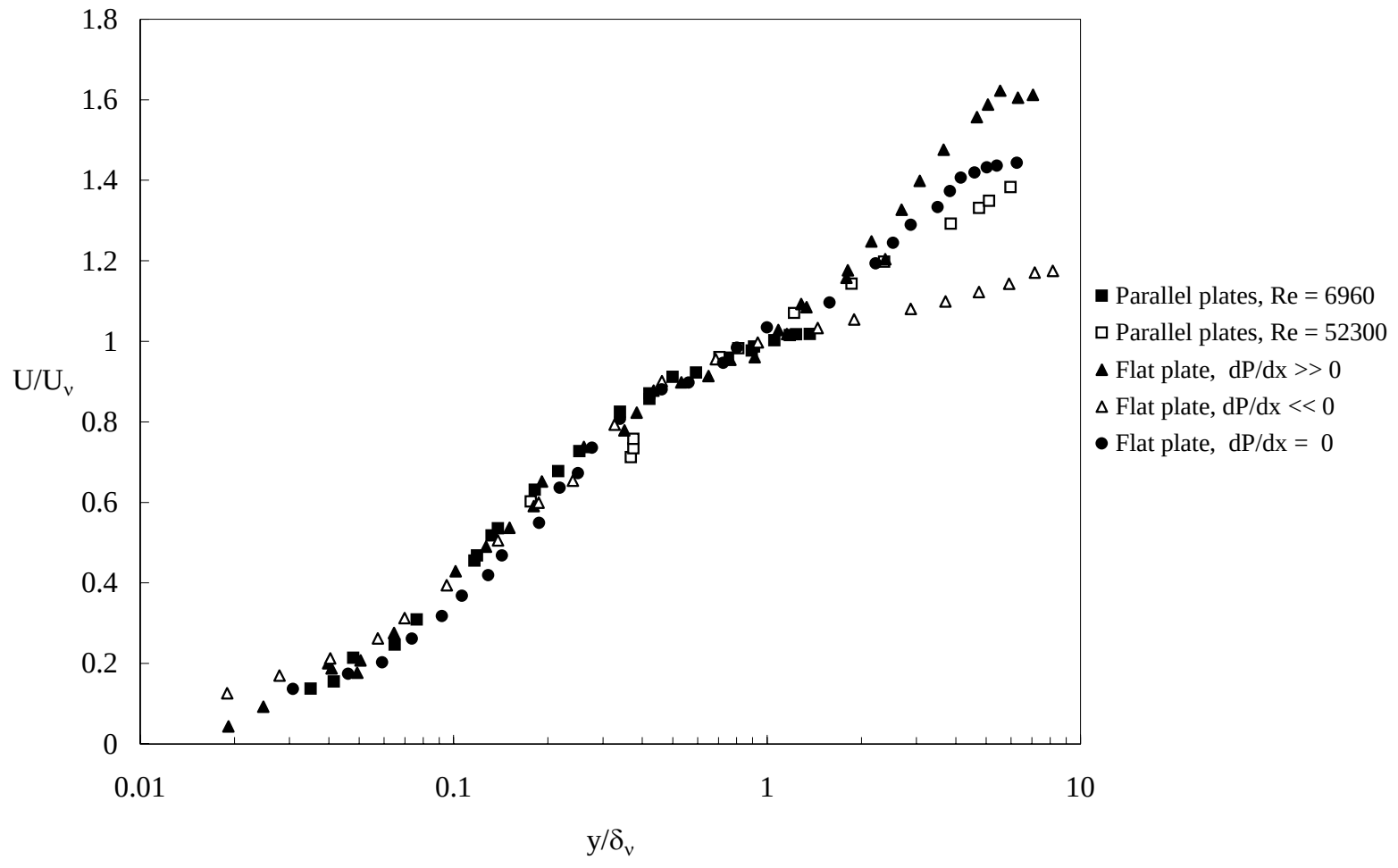

Figure 5. Zonal similarity flow near a flat surface. Data of Schlinder and Sage (1953) for parallel plates, Data of Kline et al. (1967) for boundary layers on a flat plate

Figure 5 shows that the zonal similarity analysis collapses all the velocity profiles in the inner region (wall layer and log-law) quite well into a curve which is independent of the pressure gradient above the surface but the profile in the outer region is highly dependent on the on the pressure gradient in the main flow.

## Non-Newtonian flow

It has long been known that the wall layer in non-Newtonian turbulent flow is thicker than in Newtonian fluids and results in a phenomenon called drag reduction e.g. (Elata et al., 1966). This is particularly evident in viscoelastic fluids as shown in Figure 6. Again the zonal similarity analysis collapses the data into a single curve for the inner region. The data is quite scattered, especially in the wall layer because the measurements of velocity in viscoelastic fluids very close to the wall is very difficult because the disturbances created by the probes themselves.

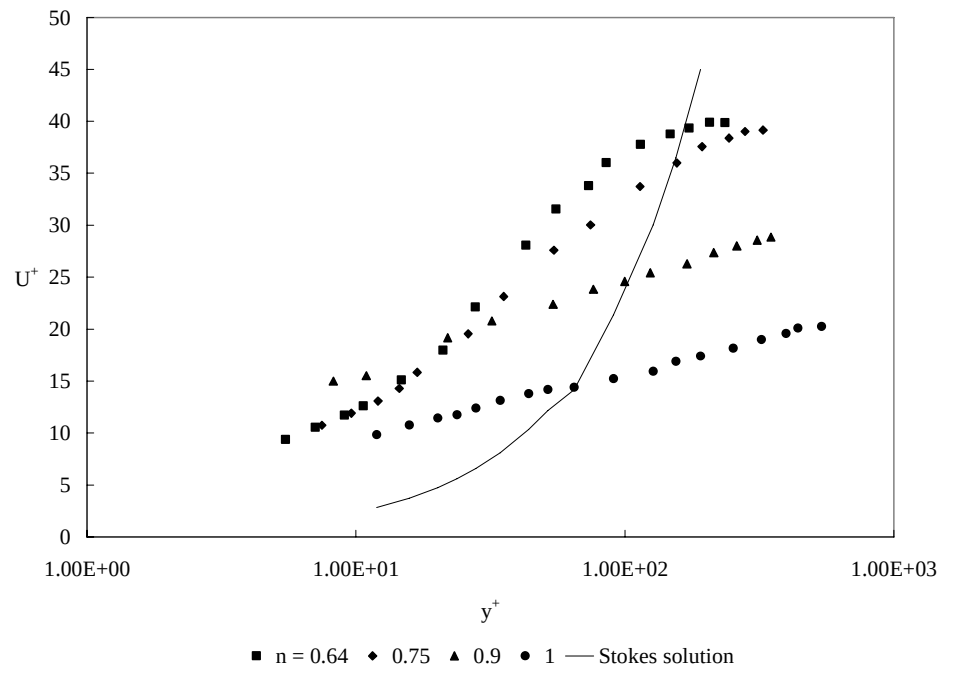

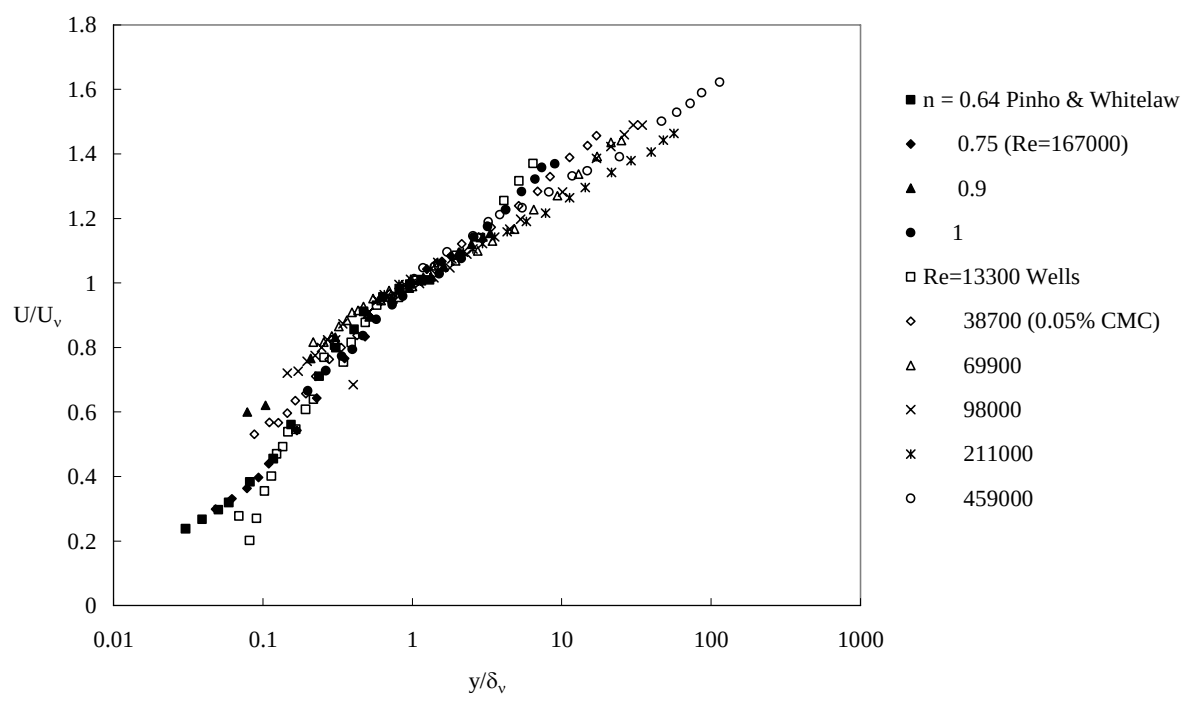

Figure 6. Zonal similar velocity profiles of different viscoelastic fluids at one Reynolds number (Pinho and Whitelaw, 1990) and of one fluid ay different Reynolds numbers (Wells, 1965).

## 3.4 Zonal similar profiles for all configurations and fluids

The profiles for all fluids and configurations studied is summarised in Figure 7. Because of crowding only one set of data for each situation has been included.

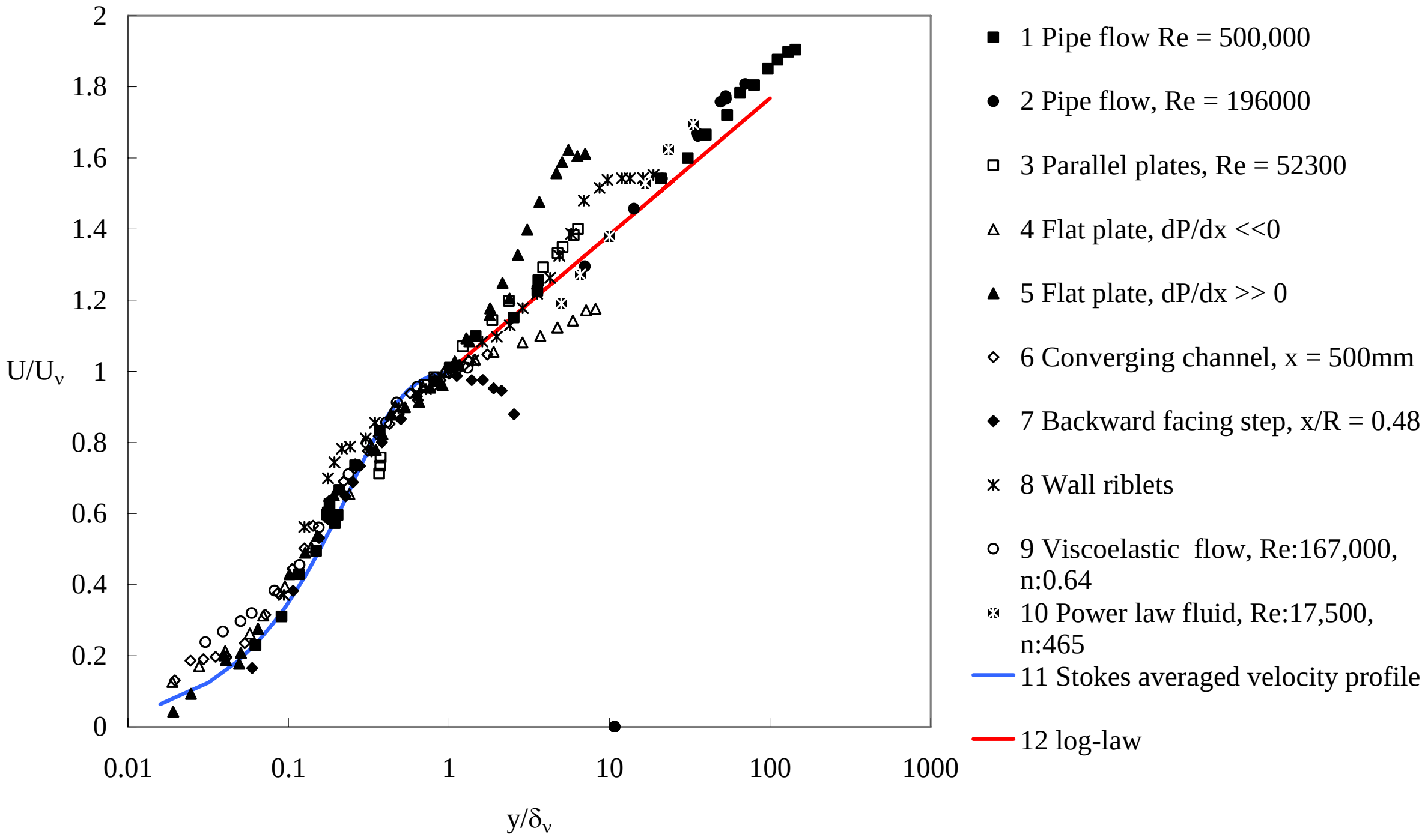

Figure 7. Zonal similar velocity profile for different types of fluids and flow configurations. Data from [1] Laufer (1954), [2, 10] Bogue (1962), [3] Schlinder & Sage (1953), [4,5] Kline *et al* (1967),[6] Tanaka & Yabuki (1986), [7] Devenport and Sutton (1991), [8] Bandhopadyay (1986), [9]Pinho and Whitelaw (1990)

The configurations covered include Newtonian (lines 1, 2), purely viscous non-Newtonian (10) and viscoelastic (9) pipe flows, flow between parallel plates (3), boundary layer with pressure gradients (4,5), converging channel flow (6) recirculating flow (7), drag reducing flow induced by wall riblets (8) and cover a wide variety of flow parameters. The zonal similarity analysis collapses all profiles in the inner region into a unique master curve. Flow in the outer region varies according to the pressure pattern and hence geometry of the main flow.

### 3.5 Thickness of the wall layer

As illustrated in Figure 3a and 4a, the intersection of the Stokes solution, equation (4) with measured velocity profiles, that defines the thickness of the wall layer, occurred at different positions from the wall depending on the Reynolds number in pipe flow or distance along the wall in a recirculation region. This thickness is also a function of the rheological behaviour of the fluid (Figure 6a). Typical examples of the wall layer thicknesses are shown in Table 1 for varying distance along the wall and Table 2 for varying fluid properties and pipe Reynolds numbers.

Table 1 Dimensionless wall layer thickness behind a backward facing step calculated from the data of Devenport and Sutton (1991).

| Distance x/R | 0.16 | 0.32 | 0.48 | 0.64 | 0.8 | 1.12 | 1.28 | 1.6 | 1.92 |
|---|---|---|---|---|---|---|---|---|---|
| Wall layer thickness $\delta_{\nu}^{+}$ | 19.8 | 36.1 | 36.1 | 28.6 | 17.7 | 31.5 | 36.1 | 45.6 | 50.4 |

Table 2 Dimensionless wall layer thickness in pipe flow for different Reynolds numbers and fluid properties

| Type of fluid | n | Re | $\delta_{\nu}^{+}$ | Source | Remarks |
|---|---|---|---|---|---|
| Newtonian | 1 | 2600 | 83.2 | Bogue (1962) | |
| | | 3300 | 75.0 | | |
| Viscous, | 0.745 | 3660 | 80.1 | | Metzner-Reed (1955) |
| non-Newtonian | 0.70 | 11700 | 80.05 | | generalised Reynolds number |
| | 0.59 | 6100 | 85.0 | | |
| | 0.53 | 17400 | 79.85 | | |
| | 0.465 | 7880 | 80.0 | | |
| Viscoelastic | 1 | 16700 | 60.0 | Pinho & Whitelaw (1990) | Reynolds number based on the non-Newtonian viscosity at the average wall shear stress |
| | 0.90 | 16700 | 105.0 | | |
| | 0.75 | 16700 | 155.0 | | |
| | 0.64 | 16700 | 180.0 | | |
| | | 459000 | 60 | Wells (1968) | Reynolds number based on the solvent viscosity |
| | | 98000 | 74 | | |
| | | 38700 | 82 | | |
| | | 211000 | 60 | | |
| | | 69900 | 75.6 | | |
| | | 13300 | 88 | | |

The wall layer thickness increases as the Reynolds number decreases and as the fluid becomes more non-Newtonian. It becomes particularly thick for viscoelastic fluids.

## 4 Discussion

The distinctive feature of this analysis is simplicity which is also its strength. Unlike other analyses such as the shift parameter models Oberlack (1999),Wosnik et al. *(*.2000), Buschmann and Gad-el-Hak (2003, 2004), McKeon et al. *(*2005) and the stress gradient balance model Fife et al. (2005), it does not require any modelling assumption; it is based on widely published experimental evidence that two zones, dominated by completely different transient coherent structures, exist in wall-bounded turbulent flow fields, the wall layer and the outer region, connected by an intermediate log-law zone. Much more experimental data has emerged since this zonal similarity was first presented to colleagues in Australasia (Trinh, 1994) after the author was allowed to emigrate, from a then-closed Stalinist regime in Viet Nam after twenty years of seclusion from the world scientific community, but examination of this new data such as the Princeton superpipe data shows that the zonal similarity profile holds and no modification to the original graphical presentation was really necessary.

Despite this simplicity, the zonal similarity analysis provides insight into many of the issues that are central to modern turbulence particularly: universality and invariance. In their review, Meneveau and Katz (2000) define scale invariance as the property by which certain features remain the same in different scales of motion. They give as example the well-known Kolmogorov universal power law spectrum

$$E(k) = c_k \varepsilon^{2/3} k^{-5/3} \qquad (6)$$

Where $c_k$ is the Kolmogorov constant, $\varepsilon$ the dissipation rate of kinetic energy by molecular viscosity and k the wavenumber. There is considerable evidence that the exponents in equation (6) hold for a range of scales that Kolmogorov called the inertial sub-range e.g. Sreenivasan (1995) but there is considerable less proof that this range of scale is the same for all flows (in terms of absolute value). By universality, Biferale et. al.(2004) mean that the small scale fluctuations are statistically independent of the large scale set-up.

The present analysis has given evidence of a unique similarity profile for the inner region when the velocity $U_\nu$ and thickness scale $\delta_\nu$ at the edge of the wall layer are used as normalising parameters. The similar velocity profile in the wall layer is unique and common to all situations considered. In that sense, it shows universality as defined by Biferale et al. but the actual values of the scales are not, as shown by the variations of $U_\nu^+$ and $\delta_\nu^+$ as shown in Tables 1 and 2.

The Stokes velocity profile given by equation (5) was time averaged and plotted in Figure 7. It matches the experimental data quite well. This supports the concept of Einstein-Li (1956) and others (Hanratty, 1956, Black, 1969, Reichardt, 1971, Meek and Baer, 1970) of using the Stokes solution to model the intermittent wall layer, especially since this solution can also give surprisingly good predictions for the probability density function of the instantaneous velocity (Trinh, 2005a) the correlation function and the "moving front of turbulence" (Trinh, 1992a) measured by Kreplin and Eckelmann (1979). But it is not the only unsteady solution that can be used to model the wall layer (Trinh, 1992a). The log-law is best expressed by forcing equation (1) through the point ($U_\nu^+$, $\delta_\nu^+$), which gives equation (7) plotted in Figure 7.

$$\frac{U^+}{U_\nu^+} = \frac{A}{U_\nu^+}\ln\left(\frac{y^+}{\delta_\nu^+}\right) + 1 \approx 2.5\ln\left(\frac{y^+}{\delta_\nu^+}\right) + 1 \qquad (7)$$

The parameter A is often expressed as the inverse of Karman's universal constant $\kappa$ but many authors have noted that this is not strictly a constant for all Reynolds numbers and configurations. Fife et al (2005) argue in their analysis that a traditional universal log law as expressed by equation (1) requires a constant value for A and that it is unlikely to be exactly true. It is shown elsewhere (Trinh, 2005b, Trinh, 1992a) that the Karman constant can be explained in structural terms as the tangent of the angle that the shear layers created by the ejections make with the normal distance $y^+$ used in most models of turbulence. This angle is likely to be a function of the relative strengths of the ejections and the main flow that will deflect them the way wind deflects a plume of chimney smoke. In that sense of course only the form of the log law has universality, the values of its parameters do not. Nonetheless, the variations of $\kappa$ appear to be relatively small.

Data from all flows considered also collapsed well in the log-law zone but the similarity analysis does not apply in the outer region where the longitudinal pressure distribution and therefore the flow geometry also affect the velocity profile. The intersection between the log-law and the outer region is also highly dependent on the Reynolds number and the flow geometry. Thus the velocity distribution in the outer region is best modelled with modern computer fluid dynamic packages but a more accurate technique must be developed to match it with the log-law.

The existence of a unique master profile for the inner region greatly simplifies this matching process and also raises questions about the need for separate wall functions for low Reynolds numbers (Launder and Sandham, 2002) and non-Newtonian fluids (Welti-Chanes et al., 2005) in computational fluid dynamics (CFD) packages. The writer believes that a better approach is to develop rules for estimating the boundaries between the inner and outer regions in these different situations.

The fact that the master similarity velocity profile applies equally to positive and negative pressure gradients, as shown in Figure 5 makes it a powerful tool in CFD's that essentially relate flow geometry to pressure gradients. A clear advantage is the applicability of the master profile to recirculation regions (Figure 4) that are very poorly modelled in present CFD's using the classical parameters of the Prandtl log-law.

The existence of a unique similarity profile for the wall region in all fluids indicates that the mechanism of turbulence production near the wall is independent of the rheological properties of the fluid. When the viscoelastic properties of the fluid are increased, the velocity profile in pipe flow has been known to depart steadily from the log-law towards a limiting curve known as Virk's asymptote (Virk et al., 1970). Some have argued that this is evidence that the mechanism of production of turbulence in non-Newtonian fluids differs from that in Newtonian fluids. The data of Pinho and Whitelaw (1990) plotted in Figures 6 and 7 indicate that, for very viscoelastic fluids, the entire flow field is defined by the wall layer. The log-law has not yet made its appearance and Virk's asymptote is simply part of the Stokes solution (Trinh, 2010) . The friction factors-Reynolds number relationship of non-Newtonian fluids has been shown to differ from their Newtonian counterpart (Dodge and Metzner, 1959). Trinh (1999 , 2009) has shown that when the classic results of the Metzner school are expressed in terms of the instantaneous wall shear stress at the point of ejection of the low speed streaks rather than the traditional time averaged shear stress, all data collapse onto the Newtonian curve. Thus clearly the difference lies in the integration constant that is implied when the Navier-Stokes equation are transformed into the Reynolds equations but never accounted for properly in semi-empirical modelling. There is no real proof that the mechanism of turbulence production is different in Newtonian and non-Newtonian fluids. Viscoelasticity does seem to dampen the velocity fluctuations imposed by the travelling hairpin vortex on the Stokes layer developed underneath its path. Consequently the ejections occur later than in purely viscous fluids and the wall layer becomes thicker as shown in table 2 but the similarity velocity profile in the wall layer remains the same.

## 5 Conclusion

A unique similarity velocity profile exists for the wall layer and log-law regions in turbulent pipe flow when the velocity profiles are normalised with the velocity and time scales at the edge of the wall layer. The method proposed here allows a quantification of the parameters at the edge of the wall layer.

The evidence indicates that the mechanism of turbulence near the wall is independent of fluid properties but the thickness of the wall layer varies significantly with flow conditions, particularly the Reynolds number, the flow geometry and the viscoelasticity of the fluid.

## 6 Nomenclature

| | |
|---|---|
| A,B | Dummy symbols |
| R | Radius |
| $y^+$ | Non-dimensional normal distance $yu_*/v$ |
| u | Local instantaneous velocity in the Stokes solution, equation (3) |
| $u_*$ | Friction velocity $u_* = \sqrt{\tau_w/\rho}$ |
| U | Time-averaged local velocity |
| $U^+$ | Non-dimensional time-averaged velocity, $U/u_*$ |
| $U_v^+$ | Non-dimensional time-averaged velocity at the edge of the wall layer |
| $\delta_v^+$ | Non-dimensional wall layer thickness |
| τ | Local shear stress |
| $\tau_{\nu}$ | Local viscous shear stress |
| $\tau_t$ | Local turbulent shear stress |
| $\tau_w$ | Wall shear stress |
| ρ | Density |
| $\nu$ | Kinematic viscosity |
| Re | Reynolds number |